\documentclass[twocolumn,secnumarabic,amssymb, nobibnotes, aps, pra,showpacs,preprintnumbers]{revtex4-1}
\usepackage{amssymb}
\usepackage{amsmath}
\usepackage{array}
\usepackage{graphicx}

\begin{document}

\title{Quantum Correlations of Helicity Entangled States in Non-inertial Frames Beyond Single Mode Approximation}%

\author{Zeynab Harsij}
\email{z.harsij@ph.iut.ac.ir}
\author{Behrouz Mirza}
\email{b.mirza@cc.iut.ac.ir}
\affiliation{Department of Physics,
Isfahan University of Technology, Isfahan 84156-83111, Iran}

\date{\today}%

\begin{abstract}
A helicity entangled tripartite state is considered in which the
degree of entanglement is preserved in non-inertial frames. It is
shown that Quantum Entanglement remains observer independent. As
another measure of quantum correlation, Quantum Discord has been
investigated. It is explicitly shown that acceleration has no
effect on the degree of quantum correlation for the bipartite and
tripartite helicity entangled states. Geometric Quantum Discord as
a Hilbert-Schmidt distance is computed for helicity entangled
states. It is shown that living in non-inertial frames does not
make any influence on this distance, either. In addition, the
analysis has been extended beyond single mode approximation to
show that acceleration does not have any impact on the quantum
features in the limit  beyond the single mode. As an interesting
result, while the density matrix  depends on the right and left
Unruh modes, the Negativity as a measure of Quantum Entanglement
remains constant. Also, Quantum Discord does not change beyond
single mode approximation.
\end{abstract}


\maketitle

\section{Introduction}\label{a}
\indent Quantum correlations are fundamental tools in Quantum
Information Theory \cite{nielsen}. They have found practical
 applications in recent discoveries in quantum cryptography
\cite{ekert,benett}, quantum teleportation \cite{bennet}, quantum
computer \cite{feynmann,shor}, and quantum dense-coding
\cite{bennett}. Up until 2001, correlations  not witnessed  by
Quantum Entanglement measures were thought not to be quantum
correlations \cite{modi}. Henderson and Vedral \cite{Henderson},
and Olivier and Zurek \cite{Zurek} introduced a new measure,
called Quantum Discord, and concluded that Quantum Entanglement
did not properly span all non-classical correlations. This measure
has been
quantitatively investigated recently \cite{leflamme2002,datta2008}.\\
\indent The study of quantum correlations in non-inertial frames
plays an important role in investigating Quantum Information in
black holes such as entropy and information paradox
\cite{bombelli,hawking}. Adesso \textit{et al} \cite{adesso} and
Mann \textit{et al} \cite{mann} showed that entanglement of a
scalar field in a non-inertial frame is affected by increasing
acceleration. Alsing and Fuentes showed degradation of Fermi-Dirac
field entanglement\cite{Alsing}. However, the degradation of
entanglement will occur from the perspective of a uniformly
accelerated observer, which essentially originates from the fact
that the event horizon appears and Unruh effect results in the
loss of information for the non-inertial observer
\cite{milburn,fuentes,Salman}. Moreover, Quantum Discord as a
measure of quantum correlation is degraded in accelerated frames
\cite{mehri}. A bipartite helicity entangled state has been
studied recently which has demonstrated that Quantum Entanglement
of such a state remains unchanged while acceleration increases
\cite{ling}. In this paper, a tripartite photon helicity entangled
is considered where, the first, second, and third observers,
respectively, represent an inertial observer,one, or two uniformly
accelerated observers. What makes this interesting is the helicity
entangled state taking the form of a mixed state, so that the
corresponding logarithmic negativity as a measure of Quantum
Entanglement remains invariant against the acceleration of the
second and third observers. This fact is completely different from
the former entangled states in which the degradation of
entanglement depended on the acceleration of the observer
\cite{Salman}. Quantum Entanglement does not include all quantum
correlations; therefore, another measure should be computed to
realize the acceleration dependence of this special correlation.
In this paper, we considered these states and investigated their
Quantum Discord as the difference between the two variant
definitions of mutual information and  showed that this quantum
feature is also observer-independent.

\indent All the observations and calculations have so far been
limited to the single mode approximation, where observers only
detect a single frequency mode \cite{Amin}. But this approximation
does not hold for all states. It is only appropriate for some
special states which form some wave-packets by imposing fourier
transformation \cite{Bruschi}. It is shown that beyond the
single-mode approximation, the Quantum Entanglement would depend
on the type of wave-packets used in accelerated systems
\cite{Bruschi1,Martin}. Studying the relativistic quantum
information beyond the single-mode approximation has been a recent
topic of investigation \cite{Friis,Ramzan}. As another basic
result, we show in this paper that this entangled state preserves
its peculiarities in the limit  beyond the single mode.

\indent The rest of the paper is organized as follows: Section 2
gives an abstract view of states in an accelerated frame. In
Section 3, Quantum Entanglement is studied and  computed for the
helicity tripartite entangled state in the non-inertial frame.
Section 4 is devoted to the measure of quantum correlations.
Quantum Discord will be computed for the specific bipartite and
tripartite entangled states to investigate whether it is observer
dependent or not. In Section 5, quantum correlations will be
analyzed beyond the single mode approximation. Here, we realize
that extending the single mode approximation would not make any
difference for the acceleration dependence of these entangled
states. Finally, the results and conclusions are presented in
Section 6.

\section{Accelerated frames}\label{b}
\begin{figure}
\includegraphics[angle=0,width=.55\textwidth]{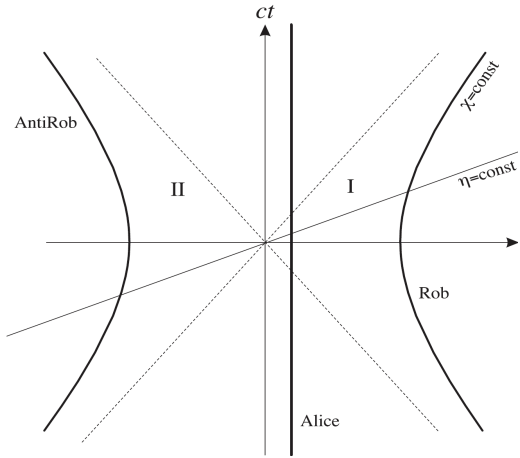}
\caption{Rindler space-time. As mentioned in the text, $ \chi , \eta
$ are Rindler coordinates and form the hyperbola. The two regions I,
II are casually disconnected. Rob would travel in either one of
these two regions.} \label{1}
\end{figure}

\indent In investigating Quantum Entanglement and Quantum Discord
in non-inertial frames, the observers should be on a hyperbola
trajectory. Therefore, we use the Rindler coordinates to define
the properties of the observers. These coordinates form four
regions in a flat space-time. Here, we work with just two regions
related to the Minkowski coordinates as follows
\begin{eqnarray}\label{b1}
 (I)\ \left\{\begin{array}{c}
              x=\frac{e^{a\chi}}{a}\cosh{a\eta} \\

              \ \ t=\ \frac{e^{a\chi}}{a}\sinh{a\eta}
           \end{array}
 \right\},\ \ \ \ (II)\ \left\{\begin{array}{c}
              x=\frac{-e^{a\chi}}{a}\cosh{a\eta} \\
                t=-\frac{ e^{a\chi}}{a}\sinh{a\eta}
           \end{array}\right\}\nonumber\\.
\end{eqnarray}
\indent In this equation, $a$ represents the uniform acceleration
of the non-inertial frame, $(x,t)$ are the Minkowski coordinates
while $(\chi,\eta)$ are the Rindler coordinates. $(I)$ is the
first Rindler region, which clearly differs from $(II)$ by its
sign while both are casually disconnected. The different signs
indicate the two different directions in time. Since no
information could flow from one region to another, we usually
trace over one of them \cite{Alsing}. The two other Rindler
regions are given by interchanging $\sinh$ by $\cosh$.
The Rindler regions are shown in Fig. \ref{1}.\\
\indent Moreover, the concepts of particle and vacuum in
non-inertial frames would change. Therefore, these are not basic
concepts and therefore are observer-dependent. This effect, known as
the Unruh effect, illustrates the fact that a vacuum in an inertial
frame would make the non-inertial observer to feel a thermal bath of
particles. In these frames the quantization is not unique and, as a
consequence, the creation and annihilation operators would be in
different forms which are related by the Bogolioubov coefficients
\cite{carol}. Bogolioubov coefficients play important roles in
defining observer dependent concepts. In this paper, we are
interested in investigating some  of the quantum properties of
entangled electromagnetic states. Consider the two observers Alice
and Bob who are going to communicate and transfer information
\cite{fuentes}, where it is formal for Alice to be inertial and Bob
to be in the accelerated frame. By this definition, Alice lives in
Minkowski space-time whereas Bob is in the Rindler space-time. Thus,
we will have
\begin{equation}\label{b2}
|\psi\rangle=|A\rangle\otimes|B\rangle,
\end{equation}
where, $|A\rangle$ is formed for Alice and $|B\rangle$ for Bob in
the accelerated frame. For Bob, the vacuum state in the Minkowski
space-time could be defined as \cite{fuentes}
\begin{equation}\label{b3}
|0\rangle^{M}= (1-e^{-2\pi\omega})^\frac{1}{2}\sum_{n=0}^{\infty}
e^{-n\pi\omega}|n\rangle^{I}|n\rangle^{II}.
\end{equation}
In this equation, we have assumed that the quantum states are
bosonic. Acceleration in $\omega$ is designated by $\frac{E}{a}$
where $a$ denotes Bob's acceleration and $E$ is the energy which
Bob detects. The other coefficients owe their presence to
normalization and to the fact that $D|0\rangle^{M}=0$, where $D$
is the Minkowski annihilation operator which could be written in
terms of creation and annihilation operators in Rindler ones:
\begin{equation}\label{b4}
D=\frac{1}{\surd{2\sinh\omega}}(e^{-\pi\omega}a^{I}-e^{\pi\omega}a^{\dagger
II}),
\end{equation}
$a^{I(II)}$, $a^{\dagger I(II)}$ are the first or the second
Rindler annihilation and creation operators, respectively. For our
purposes, the first excited state, $|1\rangle ^{M}$, is also
needed which could be easily derived by including  $D^{\dagger}$
on the vacuum state in Eq. (\ref{b3})
\begin{eqnarray}\label{b5}
D^{\dagger} |0\rangle ^{M} &=& |1\rangle ^{M} \\ \nonumber &=&
(1-e^{-2\pi\omega}) \sum_{n=0}^{\infty} e^{-2n\pi\omega}
\sqrt{(n+1)}|(n+1)\rangle^{I}|n\rangle^{II}.
\end{eqnarray}
In addition, since features of the helicity entangled states are
to be investigated, they need to be introduced in the forms below
\cite{ling}:
\begin{eqnarray} \label{b6}
|0\rangle^{M} _{\omega,p,s} &=& (1-e^{-2\pi\omega})^\frac{1}{2}
\sum_{n=0}^{\infty}
e^{-n\pi\omega}|n\rangle^{I} _{\omega,p,s}|n\rangle^{II} _{\omega,-p,s},\nonumber\\
|1\rangle ^{M}_{\omega,p,s} &=& (1-e^{-2\pi\omega})\times\\
\nonumber &\sum_{n=0}^{\infty}& e^{-n\pi\omega}
\sqrt{(n+1)}|(n+1)\rangle^{I} _{\omega,p,s} |n\rangle^{II}
_{\omega,-p,s}.
\end{eqnarray}
In the above equations, $\omega$ indicates the energy impressing
the observer, while $p$ and $s$ both show the helicity of the
photon field. Using these definitions, the helicity entangled
quantum properties can now be investigated.

\section{Quantum Entanglement}\label{c}
\indent It has been shown that maximally entangled states  have
decrement in their degree of entanglement in non-inertial frames
\cite{Alsing,fuentes}. Degradation happens when acceleration is
increased \cite{Salman,Mann}. With degradation, the information
sent by Alice to other observers would be lost, which is not
desirable. Therefore, these accelerated systems are not suitable
for transmitting information. An entangled state which seems to
send information without loss is a helicity entangled state which
is like a Werner-GHZ state with maximal entanglement but entangled
in the helicity part. In defining this state,  the helicity state
which is moving with the uniform acceleration should be defined
first. As an example, for a state which has a particle (photon)
and a positive helicity we could have
$$ |1\rangle _{+\uparrow B}^{M}= (1-e^{-2\pi\omega}) \sum_{n=0}^{\infty}
e^{-n\pi\omega}\sqrt{(n+1)}|(n+1)\rangle^{I}_{+\uparrow}|n\rangle^{II}_{-\uparrow},$$
Other states are defined in a similar fashion. Here, the subindex
$+(-)$ indicates the positive (negative) direction in momentum and
A(B) is the Alice (Bob) observer. Also, $\uparrow (\downarrow)$
shows the spin up (down) of the photon. Helicity is defined as the
multiple of spins and momentums of photon.
\\
\indent One  measure of Quantum Entanglement (QE) is Negativity
used to determine how far away two states are from separable
states \cite{negativity}. If it is equal to 1, then the state is
maximally entangled; when it goes to zero, the state is separable;
otherwise, it is partially entangled. In order to study the
negativity, we need first to calculate the partial transpose of
the density matrix which is constructed from the state given
\cite{barnett}. Peres  presented a theorem which gives a necessary
condition for separability. This theorem states that when the
eigenvalues of the partial transpose of the Alice-Bob density
matrix are positive, $\rho_{AB} ^{PT}\geq0$, then $AB$ is
separable \cite{Peres}. As already mentioned, it is just a
necessary condition and it does not guarantee the separability in
systems other than the bipartite ones. If the above condition is
satisfied, we could calculate the negativity  defined as the sum
of the negative eigenvalues of  $\rho_{AB} ^{PT}$
\cite{negativity},
\begin{equation} \label{c2}
\mathcal{N} =\log _2 \sum _{i}\mid\lambda _{i}\mid ,
\end{equation}
in which a logarithmic negativity is introduced which is commonly
used in bipartite systems. Also, we have summed over the absolute
values of the eigenvalues, $\mid\lambda _{i}\mid$. Other
definitions exist for negativity which could be used for systems
greater than bipartite ones \cite{wang}
\begin{equation} \label{c3}
\mathcal{N} =\Vert \rho ^{PT} \Vert -1,
\end{equation}
where, $\rho^{PT}$ is the partial transpose of the density matrix
in one of the subsystems of the multipartite system and $\Vert O
\Vert$ denotes the trace norm of $O$ , $Tr[ \sqrt{O^{\dagger}O}]$.
\\
\indent In order to investigate the frame dependence of QE for a
general state, we study negativity of a tripartite system in which
they are in a non-inertial frame and entangled in the helicity
part. Thus, the state we would like to study is as follows
\begin{equation}\label{c4}
|\psi\rangle =\frac{1}{\sqrt{2}} ( |1\rangle _{+\uparrow A}^{M}
|1\rangle _{-\downarrow B}^{M} |1\rangle _{-\downarrow C}^{M}+
|1\rangle _{+\downarrow A}^{M} |1\rangle _{-\uparrow B}^{M}
|1\rangle _{-\uparrow C}^{M}) .
\end{equation}
This state is similar to the Werner-GHZ one investigated in the
reference \cite{Salman}, although both are maximally entangled;
the difference, however, lies  in their types of entanglement.  In
a tripartite system, we have three subsystems; namely, Alice is an
inertial observer, Bob or Charlie, or both are the non-inertial
observers. In tripartite systems, there are two entanglement
measures: three-tangle and $\pi$-tangle \cite{ou}. The former is
not of interest in this paper (simply because it cannot be
computed analytically \cite{coffman}). The latter, however, which
is adapted as the quantification of entanglement in tripartite
systems will be investigated here. The $\pi$-tangle is defined as
the average of $\pi _{A}$, $\pi _{B}$, and $\pi _{C}$; that is:
\begin{equation} \label{c5}
\pi = \frac{1}{3} (\pi _{A}+\pi _{B} + \pi _{C}).
\end{equation}
Here, $\pi _{A}$ is defined as
\begin{equation}\label{c6}
\pi _{A} = \mathcal{N}_{A(BC)}^{2} - \mathcal{N}_{AB}^{2}
-\mathcal{N}_{AC}^{2},
\end{equation}
$\pi _{B}$ and $\pi _{C}$ are interpreted as the permutation of A,
B, and C. $ \mathcal{N}_{AB} $ is the negativity measure of the
mixed state which is $ \rho _{AB} = Tr_{C}(|\psi \rangle
_{ABC}\langle \psi |) $ and $ \mathcal{N}_{A(BC)} $ is the
one-tangle measure of negativity which is computed via
Eq.(\ref{c3}). As  will be shown in the following equations, this
measure is computed when the density matrix is transposed in the
inertial observer. This would define the entanglement between the
inertial observer and the two accelerated ones. \\
For a system in which A is inertial , B and C are non-inertial and
when we have traced over the second Rindler region, the density
matrix will be of the following form
\begin{widetext}
\begin{eqnarray}\label{c7} \rho_{AB_{I}C_{I}} &=&
Tr_{B_{II},C_{II}}(\rho_{ABC})=\frac{(1-e^{-2\pi \omega _B})^2}{2}
(1-e^{-2\pi \omega _C})^2\times \\\nonumber
&\sum_{n,m=0}^{\infty}&e^{-2n\pi\omega_B}
e^{-2m\pi\omega_C}(n+1)(m+1) \\
\nonumber &\times&(|1\rangle _{+\uparrow A} |n+1\rangle
_{-\downarrow B} |m+1\rangle _{-\downarrow C} \langle1|
_{+\uparrow A} \langle n+1|
_{-\downarrow B} \langle m+1| _{-\downarrow C}\\
\nonumber &+& |1\rangle _{+\uparrow A} |n+1\rangle _{-\downarrow
B}|m+1\rangle _{-\downarrow C}\langle1| _{+\downarrow A} \langle
n+1|
_{-\uparrow B} \langle m+1| _{-\uparrow C} \\
\nonumber &+& |1\rangle _{+\downarrow A} |n+1\rangle _{-\uparrow
B}|m+1\rangle _{-\uparrow C}\langle1| _{+\uparrow A} \langle n+1|
_{-\downarrow
B} \langle m+1| _{-\downarrow C} \\
\nonumber &+& |1\rangle _{+\downarrow A} |n+1\rangle _{-\uparrow
B}|m+1\rangle _{-\uparrow C}\langle1| _{+\downarrow A} \langle
n+1| _{-\uparrow B} \langle m+1| _{-\uparrow C}) .
\end{eqnarray}
\end{widetext}
Here, subindex I means that we have traced over the second Rindler
region in B and C. Also, $ \omega_{B}(\omega_{C}) $ shows the B
(C) acceleration frame. The partial transpose of this density
matrix on the first observer, Alice, would take the following form
\begin{widetext}\begin{eqnarray}\label{c8}
\rho_{A(B_{I}C_{I})}^{PT} &=& \frac{(1-e^{-2\pi \omega _B})^2}{2}
(1-e^{-2\pi \omega _C})^2 \\ \nonumber
&\sum_{n,m=0}^{\infty}&e^{-2n\pi\omega_B}
e^{-2m\pi\omega_C}(n+1)(m+1)
\\ \nonumber &\times&(|1\rangle _{+\uparrow A} |n+1\rangle _{-\downarrow
B} |m+1\rangle _{-\downarrow C} \langle1| _{+\uparrow A} \langle
n+1| _{-\downarrow B} \langle m+1| _{-\downarrow C} \\ \nonumber
&+& |1\rangle _{+\downarrow A} |n+1\rangle _{-\downarrow
B}|m+1\rangle _{-\downarrow C}\langle1| _{+\uparrow A} \langle
n+1| _{-\uparrow B} \langle m+1| _{-\uparrow C} \\ \nonumber
&+&|1\rangle _{+\uparrow A} |n+1\rangle _{-\uparrow B}|m+1\rangle
_{-\uparrow C}\langle1| _{+\downarrow A} \langle n+1|
_{-\downarrow B} \langle m+1| _{-\downarrow C} \\ \nonumber &+&
|1\rangle _{+\downarrow A} |n+1\rangle _{-\uparrow B}|m+1\rangle
_{-\uparrow C}\langle1| _{+\downarrow A} \langle n+1| _{-\uparrow
B} \langle m+1| _{-\uparrow C}) .
\end{eqnarray}
\end{widetext}
According to the  definition presented in Eq.(\ref{c3}), $
\mathcal{N}_{A(B_{I}C_{I})} $, i.e. negativity, will be as
follows:
\begin{eqnarray} \label{c9}
&\mathcal{N}_{A(B_{I}C_{I})} = \frac{(1-e^{-2\pi \omega _B})^2}{2}
(1-e^{-2\pi \omega _C})^2&
\\ \nonumber &\times\left(
\sum_{n,m=0}^{\infty}e^{-2n\pi\omega_B}e^{-2m\pi\omega_C} 4(n+1)
(m+1)\right) -1=1&.
\end{eqnarray}
As already mentioned above, $ \rho_{AB_I} $  is derived by tracing
over the third subsystem C
\begin{eqnarray}\label{c10}
\rho_{AB_I}&=&Tr_C (\rho_{A(B_I C_I)})=\frac{(1-e^{-2\pi \omega
_B})^2}{2} \sum_{n=0}^{\infty}e^{-2n\pi\omega_B} \nonumber\\
&\times&(n+1)(|1\rangle _{+\uparrow A} |n+1\rangle _{-\downarrow
B}\langle1| _{+\uparrow A} \langle n+1| _{-\downarrow B} \nonumber \\
 &+& |1\rangle _{+\downarrow A} |n+1\rangle _{-\uparrow
B}\langle1| _{+\downarrow A} \langle n+1| _{-\uparrow B}).
\end{eqnarray}
Based on $|1\rangle _{+\uparrow A} |n+1\rangle _{-\uparrow B}, \ \
|1\rangle _{+\uparrow A} |n+1\rangle _{-\downarrow B},\  \
|1\rangle _{+\downarrow A} |n+1\rangle _{-\uparrow B},\  \
|1\rangle _{+\downarrow A} |n+1\rangle _{-\downarrow B}$, the
matrix form of the density matrix $\rho_{AB_I}$ will be in the
following form:
\begin{equation}\label{c11}
\rho_{AB_I}=\frac{(1-e^{-2\pi \omega _B})^2}{2}
\sum_{n=0}^{\infty}e^{-2n\pi\omega_B}(n+1) \left(
  \begin{array}{cccc}
    0 & 0 & 0 & 0 \\
    0 & 1 & 0 & 0 \\
    0 & 0 & 1 & 0 \\
    0 & 0 & 0 & 0 \\
  \end{array}
\right)
\end{equation}
Clearly, this density matrix is in a diagonal form; therefore, by
getting a partial transpose on either the A or  B part, it will
not change and its eigenvalues will still be positive. Thus, as
also mentioned in the Peres theorem, such a density matrix should
not make an entangled state. Therefore, we will have $
\mathcal{N}_{AB_{I}} =0 $. $ \rho_{AC_I} $ could be derived in a
similar manner and, as a consequence, $ \mathcal{N}_{AC_{I}} =0 $.
Therefore, from Eq. (\ref{c6}), we will have
$$ \pi_A =1.$$
$ \pi_B$ could be shown to be equal to  $ \pi_C $ which is itself
equal to $1 $. Now, we could derive $ \pi $-tangle via
Eq.(\ref{c5}) as follows:
\begin{equation}\label{c12}
\pi = \frac{1}{3}(\mathcal{N}_{A(BC)}^{2}+\mathcal{N}_{B(AC)}^2 +
\mathcal{N}_{C(AB)}^2) = 1.
\end{equation}
It is evident that the tripartite system has preserved its degree
of entanglement in a non-inertial frame. This particular helicity
entangled state seems to have some other interesting consequences
which are to be investigated in the next section.

\section{Quantum Discord} \label{d}
\indent QE  introduced in the previous section is an evidence of
quantum correlations, but it does not guarantee to include all
quantum correlations. Although separable quantum states form a
kind of quantum correlation, they are not included in QE measures.
Quantum Discord (QD) is a more general evidence  first presented
by Olivier and Zurek \cite{Zurek}. This quantity defines the
degree of quantum correlations and is defined as the difference
between two expressions of mutual information in quantum while
they are considered to be identical in classical terms
\cite{mutual}. In classical information theory, mutual information
is the correlation between random variables and  takes the
following form for a bipartite system \cite{Zurek}
\begin{equation}\label{d1}
\mathcal{J}(\mathcal{X},\mathcal{Y})=\mathcal{H}(\mathcal{X})-
\mathcal{H}(\mathcal{X}|\mathcal{Y}),
\end{equation}
where, $\mathcal{H}$ is the Shanon entropy and is given by $
\mathcal{H}=-\sum \mathcal{P}(\mathcal{X}=x)\log
\mathcal{P}(\mathcal{X}=x) $. Here, $ \mathcal{P}(\mathcal{X}) $
is the probability distribution for the random variable $
\mathcal{X} $ to have the $x$-value. $ \mathcal{H}
(\mathcal{X}|\mathcal{Y}) $ is the conditional entropy and may be
written as:
\begin{equation}\label{d2}
\mathcal{H}(\mathcal{X},\mathcal{Y}) -\mathcal{H}(\mathcal{Y}),
\end{equation}
where, $ \mathcal{H}(\mathcal{X},\mathcal{Y}) $ is the joint
entropy; i.e., both $ \mathcal{X} $ and $ \mathcal{Y} $ occurring.
Another expression for mutual information could be written in the
following form\cite{barnett}
\begin{equation}\label{d3}
\mathcal{I}(\mathcal{X};\mathcal{Y})=\mathcal{H}(\mathcal{X})+
\mathcal{H}(\mathcal{Y})-\mathcal{H}(\mathcal{X},\mathcal{Y}).
\end{equation}
It is evident that the two Eqs.(\ref{d1}) and (\ref{d3}) are
equivalent in classical theory but they will behave differently
when they are extended into the quantum systems. The difference
would lie in the term
 QD. In the quantum version, $ \mathcal{H} $ would explain the
Von-Neumann entropy $ \mathcal{S} $ which is defined in terms of
density matrix as
\begin{equation}\label{d4}
\mathcal{S}=-Tr_{\mathcal{X}} \rho_{\mathcal{X}} \log _{2}
\rho_{\mathcal{X}}.
\end{equation}
Thus, for a bipartite system, Eq.(\ref{d3}) would take the form
below:
\begin{eqnarray}\label{d5}
\mathcal{I}(\mathcal{X};\mathcal{Y})&=&\mathcal{S}(\mathcal{X})+
\mathcal{S}(\mathcal{Y})-\mathcal{S}(\mathcal{X},\mathcal{Y})\\
\nonumber &=& -Tr(\rho_{\mathcal{X}}\log _2 \rho_{\mathcal{X}})
-Tr (\rho_{\mathcal{Y}}\log _2 \rho_{\mathcal{Y}})\\
\nonumber &+& Tr(\rho_{\mathcal{XY}}\log _2 \rho_{\mathcal{XY}}).
\end{eqnarray}
In addition,  Eq. (\ref{d1}) would also change in the quantum
system. Since the conditional entropy requires the state of $
\mathcal{X} $ to be in a given state of A, we need an optimized
measurement approach \cite{barnett}. This will be achieved by
introducing some projection operators. Applying the optimized
measurement approach would change Eq. (\ref{d1}) into the
following form \cite{mehri,Ali}
\begin{equation}\label{d6}
\mathcal{J}(\mathcal{X};A)=\mathcal{S}(\mathcal{X})-min_{\pi_{i}}
[\mathcal{S}(\rho_{\mathcal{X}|\pi_{i}^A})].
\end{equation}
Evidently, the new expression for $ \mathcal{J} $ differs from the
former in their second term. This term is the optimized
measurement of state $ \mathcal{X} $ corresponding to $ \pi_{i}^A
$. The state is given as \cite{luo}
\begin{equation}\label{d7}
\rho_{\mathcal{X}|\pi_{i}^A}=\frac{1}{P_i} \pi_{i}^A
\rho_{\mathcal{X},A}\pi_{i}^A,
\end{equation}
where, $ P_i $ is equal to $ Tr_{\mathcal{X},A} (\pi_i ^A
\rho_{\mathcal{X},A}) $. It is the probability for each
measurement to have a given value. In the forthcoming subsection,
we will present an explicit expression for QD in a bipartite
system and will compute it for a particular case.

\subsection{Quantum Discord for a bipartite helicity entangled state}\label{da}
\begin{figure}
\includegraphics[angle=0,width=.5\textwidth]{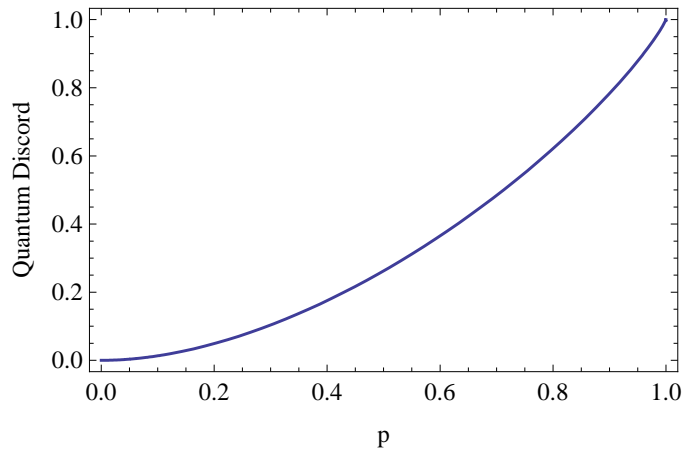}
\caption{Quantum Discord for a bipartite system versus p explicitly
shows only p-dependence. } \label{2}
\end{figure}

\begin{figure}
\includegraphics[angle=0,width=.5\textwidth]{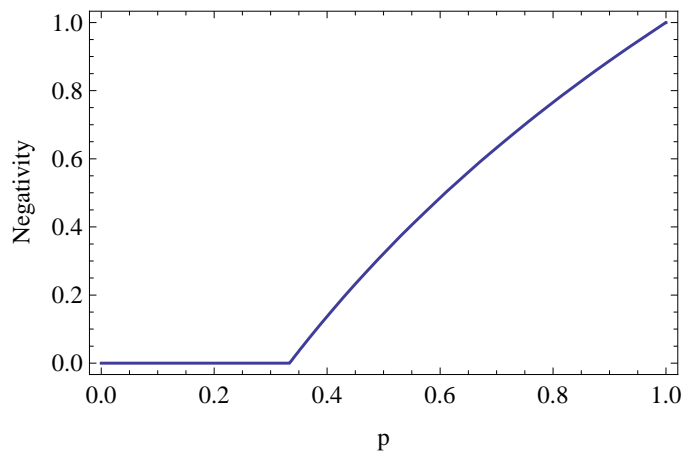}
\caption{Negativity versus p. For $ p\geq \frac{1}{3} $, the state
is entangled. No observer (acceleration) dependence is seen.}
\label{3}
\end{figure}

\indent In the previous section, we investigated Quantum
Entanglement for an entangled state. It was shown  that this
particular state does not behave as a usual entangled state does
when it is observed in an accelerated frame. Similar to QE, QD is
degraded by increasing acceleration \cite{mehri}. Here, we would
like to determine whether this specific case has similar features
in QD. A density matrix which is commonly used for computing QD
for a two-state system is as follows \cite{barnett}
\begin{equation}\label{d8}
\rho=(\frac{1-p}{4})I+p|\psi\rangle \langle \psi|.
\end{equation}
In this density matrix, $ I $ is the identity and p is the
probability given for finding a state in one of the states $
|0\rangle $ or $ |1\rangle $. The probability is, therefore,
bounded, $ 0\leq p \leq 1 $. When $ p=1 $, $ \rho $ is a pure
state and when $ p=0 $, it is the identity. Using the following
bipartite entangled state
\begin{equation}\label{d91}
|\psi\rangle =\frac{1}{\sqrt{2}} ( |1\rangle _{+\uparrow A}^{M}
|1\rangle _{-\downarrow B}^{M} + |1\rangle _{+\downarrow A}^{M}
|1\rangle _{-\uparrow B}^{M}),
\end{equation}
we can determine  QD for a helicity entangled state. By putting
the state given in Eq.(\ref{d91}) into Eq.(\ref{d8}) and by
tracing over the second Rindler region for the Bob observer, we
will clearly have the following density matrix
\begin{eqnarray}\label{d9}
\rho_{AB_I}&=&(1-e^{-2\pi\omega})^{2}\sum_{n=0}^{\infty}
e^{-2n\pi\omega}(n+1) \times\\ \nonumber
&(&\frac{1-p}{4}(|1\rangle_{+\uparrow A}|n+1\rangle_{-\uparrow
B}\langle1|_{+\uparrow A}\langle n+1|_{-\uparrow B}\\ \nonumber
&+& |1\rangle_{+\downarrow A}|n+1\rangle_{-\downarrow B}\langle
1|_{+\downarrow A}\langle n+1|_{-\downarrow B})\\ \nonumber &+&
\frac{1+p}{4}(|1\rangle_{+\uparrow A}|n+1\rangle_{-\downarrow
B}\langle 1|_{+\uparrow A}\langle n+1|_{-\downarrow B} \\
\nonumber &+& |1\rangle_{+\downarrow A}|n+1\rangle_{-\uparrow
B}\langle 1|_{+\downarrow A}\langle n+1|_{-\uparrow B})\\
\nonumber &+& \frac{p}{2}(|1\rangle_{+\uparrow
A}|n+1\rangle_{-\downarrow B}|1\langle_{+\downarrow A}\langle
n+1|_{-\uparrow B}\\ \nonumber &+& |1\rangle_{+\downarrow
A}|n+1\rangle_{-\uparrow B}\langle 1|_{+\uparrow A}\langle
n+1|_{-\downarrow B})).
\end{eqnarray}
Based on $|1\rangle _{+\uparrow A} |n+1\rangle _{-\uparrow B},\ \
|1\rangle _{+\uparrow A} |n+1\rangle _{-\downarrow B},\ \
|1\rangle _{+\downarrow A} |n+1\rangle _{-\uparrow B},\ \
|1\rangle _{+\downarrow A} |n+1\rangle _{-\downarrow B}$,
$\rho_{AB_I}$ could be written in the matrix form as follows:
\begin{eqnarray}
\rho_{AB_I}&=&(1-e^{-2\pi\omega})^{2}\sum_{n=0}^{\infty}e^{-2n\pi\omega}
(n+1)\\ \nonumber &\times&\left(
  \begin{array}{cccc}
   \frac{1-p}{4} & 0 & 0 & 0 \\
    0 & \frac{1+p}{4} & \frac{p}{2} & 0 \\
    0 & \frac{p}{2} & \frac{1+p}{4} & 0 \\
    0 & 0 & 0 & \frac{1-p}{4} \\
  \end{array}
\right)
\end{eqnarray}
which is an X-shaped symmetric density matrix with real matrix
elements, the difference lying in $\rho_{14}=\rho_{41}=0$. It is
now possible to express QD more quantitatively. As  mentioned
before, QD, i.e. the difference between two mutual information
pieces, takes the form below \cite{Zurek}:
\begin{eqnarray}\label{d10}
\mathcal{D}(A:B)&=&\mathcal{I}(A:B)-\mathcal{J}(A:B)\\ \nonumber
&=&\mathcal{S}(A)+\mathcal{S}(B)-\mathcal{S}(A,B)-\mathcal{S}(A)
\\ \nonumber &+& min_{\left\lbrace\pi_{i}^{A}\right\rbrace} \mathcal{S}(A|B).
\end{eqnarray}
$ \mathcal{S}(A) $, $ \mathcal{S}(B) $, and $ \mathcal{S}(A,B) $
have been already explained above. The last term is the optimized,
measured conditional entropy \cite{Zurek,barnett}. This quantity
has been presented for two qubit states in \cite{luo}. For a real
symmetric X-state, $ \mathcal{S}(A|B) $ over all projection
operators is written as \cite{Ali}:
\begin{equation}\label{d11}
\mathcal{S}_{\pi_{i}}(A|B)=p_0\mathcal{S}(\rho_{A|0})+p_1
\mathcal{S}(\rho_{A|1}).
\end{equation}
In two qubit X-states, we only have two p-probabilities, $p_0$ and
$p_1$.  Probabilities are related to each other by
\begin{equation}\label{d12}
p_0=1-p_1=(\rho_{22}+\rho_{44})l+(\rho_{11}+\rho_{33})k.
\end{equation}
It should also be noticed that $ \rho_{nm}=\langle n|\rho|m\rangle
$. In addition, we have
\begin{equation}\label{d13}
\mathcal{S}(\rho_{A|j})=-\sum_{\pm}\lambda_{\pm}(\rho_{A|j})\log_2
\lambda_{\pm}(\rho_{A|j}),
\end{equation}
where, $ \lambda_{\pm}(\rho_{A|j}) $ could be expressed as $
\frac{1}{2}(1\pm \theta_j) $, and $ \theta_j $s are
\begin{eqnarray}\label{d14}
\theta
_0&=&\frac{1}{p_0}\sqrt{((\rho_{11}-\rho_{33})k+(\rho_{22}-\rho_{44})l)^2
+\beta}  \\ \nonumber \theta
_1&=&\frac{1}{p_1}\sqrt{((\rho_{11}-\rho_{33})l+(\rho_{22}-\rho_{44})k)^2
+\beta},
\end{eqnarray}
and $ \beta =4kl(\rho_{14}+\rho_{23})^2 -16\mu\rho_{14}\rho_{23}
$. It is shown in \cite{Ali} that the minimum value occurs when $
k=l=\frac{1}{2} $, $ \mu=0 $ or $ k=1-l=0,1 $ . According to the
above equations, QD could be computed for any two qubit states
\begin{equation}\label{d15}
\mathcal{D}(A|B)=\mathcal{S}(B)-\mathcal{S}(A,B)+min_{\pi_i}[\mathcal{S}(A|B)]
\end{equation}
For the specific state  defined in  Eq. (\ref{d91}),
$$\mathcal{S}(B)=-\sum \lambda_{\rho_B} \log_2 \lambda_{\rho_B},$$
where $\rho_B=Tr_A(\rho_{AB_I})=\frac{1}{2}I,$ which is a multiple
of identity. Thus, $\mathcal{S}(B)=1$. $\mathcal{S}(A,B)$ is
defined in terms of the eigenvalues of $\rho_{AB_I}$. The
eigenvalues are given by
$$\lambda_1=\lambda_2=\lambda_3=(1-e^{-2\pi\omega})^{2}\sum_{n=0}^
{\infty}e^{-2n\pi\omega}(n+1)\frac{1-p}{4},$$
$$\lambda_4=(1-e^
{-2\pi\omega})^{2}\sum_{n=0}^{\infty}e^{-2n\pi\omega}(n+1)\frac{1+3p}{4}.$$
Therefore, $\mathcal{S}(A,B)$ is in the following form
\begin{eqnarray}
\mathcal{S}(A,B)&=&-Tr\left(\rho_{AB}\log_2 \rho_{AB}\right) \\
\nonumber
&=&-3(\frac{1-p}{4})\log_2(\frac{1-p}{4})-(\frac{1+3p}{4})\log_2
(\frac{1+3p}{4}).
\end{eqnarray}
\indent The last term which quantifies the QD could be computed
from  Eq. (\ref{d13}). For the particular state given in Eq.
(\ref{d91}), $\lambda_\pm =\frac{1}{2}(1\pm\frac{p}{2})$.
Therefore, via Eq. (\ref{d14}), QD takes the following form
\begin{eqnarray}\label{d16}
\mathcal{D}(A|B)&=& 1+3(\frac{1-p}{4})\log _2 (\frac{1-p}{4}) +
(\frac{1+3p}{4})\log _2 (\frac{1+3p}{4}) \nonumber
\\ &-&\frac{1}{2}(1+\frac{p}{2})\log _2 \frac{1}{2}(1+\frac{p}{2})
-\frac{1}{2}(1-\frac{p}{2})\log _2 \frac{1}{2}(1-\frac{p}{2})
\nonumber
\\ &=&\frac{1}{4}\log_2 \frac{(1+3p)^{1+3p}(1-p)^{1-p}}{(1+p)^{2(1+p)}}.
\end{eqnarray}
Clearly, the quantity  acceleration dependence disappears. We
expect QD to be degraded by increasing acceleration whereas the
above term does not depend on acceleration. Fig. \ref{2}
 illustrates how QD  depends on the p-parameter and  is
independent of acceleration. We could also study negativity for
this density matrix using  the same method introduced in Section
\ref{b}. The result is shown in Fig. \ref{3}. It is observed that
negativity for such a helicity entangled state is independent  of
acceleration, as it was in previous cases. It is interesting to
note in Fig. \ref{3}  that if $ p\geq\frac{1}{3} $, then an
entangled state results.

\subsection{Global Quantum Discord}\label{db}
\begin{figure}
\includegraphics[angle=0,width=.5\textwidth]{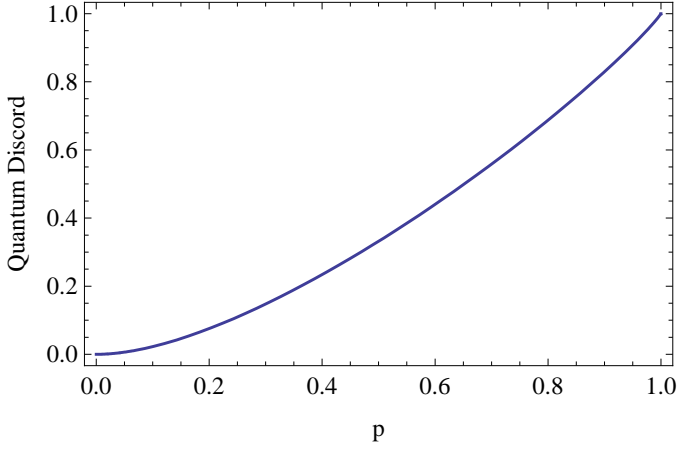}
\caption{Quantum Discord for a tripartite system versus p.
Acceleration has no influence and this Figure exhibits a milder
curvature than  Fig. \ref{2}.}\label{4}
\end{figure}

It is now desirable to extend the previous expression for QD to
multipartite systems. For any arbitrary multipartite state, there
is a global definition of QD which holds for any set of projection
measurements $ \left\lbrace \pi_{i} \right\rbrace $ \cite{Rulli}:
\begin{eqnarray}\label{d17}
\mathcal{D}(\rho_{A_1} ... \rho_{A_N})&=&min[\mathcal{S}(\rho_{A_1
... A_N}||\phi (\rho_{A_1 ... A_N}) ) ] \\
\nonumber &-&\sum_{j=1}^N \mathcal{S}(\rho_{A_j}||\phi
(\rho_{A_j}) ),
\end{eqnarray}
where, $ \mathcal{S}(\rho||\phi (\rho) )
=\mathcal{S}(\rho)-\mathcal{S}(\phi (\rho) )$ is the relative
entropy \cite{barnett}, $\phi_j (\rho_{A_j})$ is equal to $
\sum_{i} \pi_{A_j}  ^i \rho_{A_j}\pi_{A_j}  ^i $ and $ \rho_{A_j}
$s are the density matrices for the different parts. Also, $
\phi(\rho) $ could be found as $ \sum_{k} \pi_k \rho_{A_1 ...A_N}
\pi_k $ with $ \pi_k =\pi_{A_1}^{j_1}\otimes ... \otimes
\pi_{A_N}^{j_N} $, which is a set of projection measurements made.
This Global definition of QD satisfies the features of QD as
stated in \cite{Rulli}. In addition to the generalization made for
QD, we would like to extend the density matrix given in Subsection
\ref{da} Eq.(\ref{d8}) to more than a bipartite state. For a
multipartite system, the density matrix Eq.(\ref{d8}) would be
extended as follows
\begin{equation}\label{d18}
\rho=\frac{1-p}{2^N}I+p|\psi\rangle \langle \psi|.
\end{equation}
In the above expression, $I$ is a $2^N$ identity matrix and p is
the probability as defined before. When $ p\neq 0,1 $,
Eq.(\ref{d16}) provides a fully correlated state. It seems that
the approach  to be used here is different from the one used in
the last subsection, but
one could do the same calculations for $N=2$ to obtain similar  results.\\
\indent Let us now illustrate  the global definition of QD with an
example. The simplest example for a system higher than the
bipartite one is a tripartite system, the density matrix for which
is  as follows:
\begin{equation}\label{d19}
\rho_{ABC}=\frac{1-p}{8}I+p|\psi\rangle \langle \psi|.
\end{equation}
Here, $ |\psi\rangle $ is exactly the state introduced in
Eq.(\ref{c4}). Therefore, by tracing over the second regions, the
density matrix would be of the following form
\begin{widetext}
\begin{eqnarray}\label{d20} \rho_{AB_IC_I}&=&
\frac{(1-e^{-2\pi \omega _B})^2}{2} (1-e^{-2\pi \omega _C})^2 \\
\nonumber &\sum_{n,m=0}^{\infty}&e^{-2n\pi\omega_B}
e^{-2m\pi\omega_C}(n+1)(m+1)((\frac{1-p}{8}) \times \\
\nonumber &(&|1\rangle _{+\uparrow A} |n+1\rangle _{-\uparrow B}
|m+1\rangle _{-\uparrow C} \langle1| _{+\uparrow A} \langle n+1|
_{-\uparrow B} \langle m+1| _{-\uparrow C} \\ \nonumber &+&
|1\rangle _{+\uparrow A} |n+1\rangle _{-\uparrow B} |m+1\rangle
_{-\downarrow C} \langle1| _{+\uparrow A} \langle n+1| _{-\uparrow
B} \langle m+1| _{-\downarrow C}\\ \nonumber &+& |1\rangle
_{+\uparrow A} |n+1\rangle _{-\downarrow B} |m+1\rangle
_{-\uparrow C} \langle1| _{+\uparrow A} \langle n+1| _{-\downarrow
B} \langle m+1| _{-\uparrow C} \\ \nonumber &+& |1\rangle
_{+\downarrow A} |n+1\rangle _{-\uparrow B} |m+1\rangle
_{-\downarrow C} \langle1| _{+\downarrow A} \langle n+1|
_{-\uparrow B} \langle m+1| _{-\downarrow C}\\ \nonumber &+&
|1\rangle _{+\downarrow A} |n+1\rangle _{-\downarrow B}
|m+1\rangle _{-\uparrow C} \langle1| _{+\downarrow A} \langle n+1|
_{-\downarrow B} \langle m+1| _{-\uparrow C} \\ \nonumber
&+&|1\rangle _{+\downarrow A} |n+1\rangle _{-\downarrow B}
|m+1\rangle _{-\uparrow C} \langle1| _{+\downarrow A} \langle n+1|
_{-\downarrow B} \langle m+1| _{-\downarrow C})\\ \nonumber
+&(\frac{1+3p}{8})&(|1\rangle _{+\uparrow A}|n+1\rangle
_{-\downarrow B} |m+1\rangle_{-\downarrow C} \langle1| _{+\uparrow
A} \langle n+1| _{-\downarrow B} \langle m+1| _{-\downarrow C}\\
\nonumber &+& |1\rangle _{+\downarrow A} |n+1\rangle _{-\uparrow
B} |m+1\rangle _{-\uparrow C} \langle1|
_{+\downarrow A} \langle n+1| _{-\uparrow B} \langle m+1| _{-\uparrow C})\\
\nonumber +&\frac{p}{2}& (|1\rangle _{+\uparrow A} |n+1\rangle
_{-\downarrow B} |m+1\rangle_{-\downarrow C} \langle1|
_{+\downarrow A} \langle n+1| _{-\uparrow B} \langle m+1|
_{-\uparrow C}\\ \nonumber&+& |1\rangle _{+\downarrow A}
|n+1\rangle _{-\uparrow B} |m+1\rangle _{-\uparrow C} \langle1|
_{+\uparrow A} \langle n+1| _{-\downarrow B} \langle m+1|
_{-\downarrow C})).
\end{eqnarray}
\end{widetext}
Prior to the computation process, the measurements need to be
defined first. If rotations are considered  as projection
operators, then in the directions of the basis vectors of $ A,B,$
and $C $ they may be defined as
\begin{eqnarray}\label{d21}
|+\rangle _j&=&\cos(\frac{\theta _j}{2})|\uparrow\rangle_j +
e^{\imath\phi_j}\sin(\frac{\theta_j}{2})|\downarrow\rangle_j,\\
\nonumber |-\rangle _j&=&-e^{-\imath\phi_j}\sin(\frac{\theta
_j}{2})|\uparrow\rangle_j
+\cos(\frac{\theta_j}{2})|\downarrow\rangle_j,
\end{eqnarray}
where $ j=A,B,C $. It should be noted that $ \theta_j
\epsilon\left[ 0, \pi \right) $, $ \phi_j \epsilon\left[ 0, 2\pi
\right) $ and the projection operators are found to be $
\pi_{A_j}=|\pm\rangle\langle\pm| $ . Using the global definition,
Eq.(\ref{d17}), and inserting $N=3$, we get the following equation
for QD in tripartite states
\begin{eqnarray}\label{d22}
\mathcal{D}(\rho_{ABC})&=&min[
\mathcal{S}(\rho||\phi(\rho))-\mathcal{S}(\rho_A||\phi_A(\rho_A))\\
\nonumber
&-&\mathcal{S}(\rho_B||\phi_B(\rho_B))-\mathcal{S}(\rho_C||\phi_C(\rho_C))].
\end{eqnarray}
For this specific tripartite state, we have:
$$\phi(\rho_{ABC})=\sum_{k=\pm} \pi_k \rho_{ABC}\pi_k,$$
$$\pi_k= \pi_A\otimes \pi_B\otimes \pi_c.$$
By tracing over the two subsystems B and C, $ \rho_A $ is simply
derived. Since $ \rho_A $ is somehow a multiple of identity,
therefore, $ \mathcal{S} (\rho||\phi(\rho))=0 $. Similar
expressions
 hold for $ \rho_A $ and $ \rho_c $. Hence, we could simply have
\begin{equation}\label{d23}
\mathcal{D}(\rho_{ABC})=min_{\theta_j, \phi_j}\left[ \mathcal{S}
(\rho||\phi(\rho))\right].
\end{equation}
$ \mathcal{S}(\rho) $ and $ \mathcal{S}(\phi(\rho)) $ are known by
their eigenvalues as $-Tr(\rho\log_{2}\rho)$; hence,  they should
be identified. As there are many parameters to be defined, we
should like to make it  simpler by considering the two cases of  $
\theta_1=0 $ and $ \phi_j=0 $ \cite{Rulli}. The first term in
$\rho_{ABC}$, Eq.(\ref{d19}), is a multiple of  identity. It will,
therefore, suffice to minimize the second part of the density
matrix
\begin{equation}\label{d24}
\mathcal{D}(\theta_2, \theta_3)=min[-\sum_j \lambda_j\log
\lambda_j],
\end{equation}
where, $ \lambda_j $s are the eigenvalues of $
\mathcal{S}(\rho||\phi(\rho)) $ \cite{Rulli}
\begin{eqnarray}\label{d25}
\lambda_1 &=&\lambda_8 =\frac{1}{2}\cos ^2(\frac{\theta_2}{2})\cos
^2(\frac{\theta_3}{2}),\\ \nonumber \lambda_2 &=&\lambda_7
=\frac{1}{2}\cos ^2(\frac{\theta_2}{2})\sin
^2(\frac{\theta_3}{2}),\\ \nonumber \lambda_3 &=&\lambda_6
=\frac{1}{2}\sin ^2(\frac{\theta_2}{2})\cos
^2(\frac{\theta_3}{2}),\\ \nonumber \lambda_4 &=&\lambda_5
=\frac{1}{2}\sin ^2(\frac{\theta_2}{2})\sin
^2(\frac{\theta_3}{2}).
\end{eqnarray}
It is clearly seen that minimization occurs when $
\theta_2=\theta_3=0 $. Now, the Expression for QD will take the
following form
\begin{eqnarray}\label{d26}
\mathcal{D}(\rho_{ABC})&=& [\mathcal{S}(\rho||\phi(\rho))]_{\theta_j=\phi_j=0}\\
\nonumber &=&7(\frac{1-p}{8}) \log_2 (\frac{1-p}{8}) + (\frac{1 +
7p}{8}) \log_2(\frac{1 + 7p}{8})  \\ \nonumber
&-&6(\frac{1-p}{8})\log_2 (\frac{1-p}{8}) - 2(\frac{1+3p}{8})
\log_2 (\frac{1+3p}{8})\\ \nonumber &=&\frac{1}{8}\log_2
\frac{(1+7p)^{1+7p}(1-p)^{1-p}}{(1+3p)^{2(1+3p)}}.
\end{eqnarray}
Evidently, no effect of acceleration can be observed in this
expression. The result is the same as that with the bipartite one:
i.e., simply a function of p. This dependence on p is depicted in
Fig. \ref{4}. Compared with the bipartite model (Fig. \ref{2}), it
is seen that both figures are parabola but it is more curved in
the bipartite system than it is in the tripartite one. Thus,  we
have shown another interesting feature of the particular helicity
entangled state, which contradicts the degradation of QD in a
non-inertial frame. In the next subsection, we would like to
investigate another feature of this system which has come to be
known as Geometric Quantum Discord.

\subsection{Geometric Quantum Discord}\label{dc}
\begin{figure}
\includegraphics[angle=0,width=.5\textwidth]{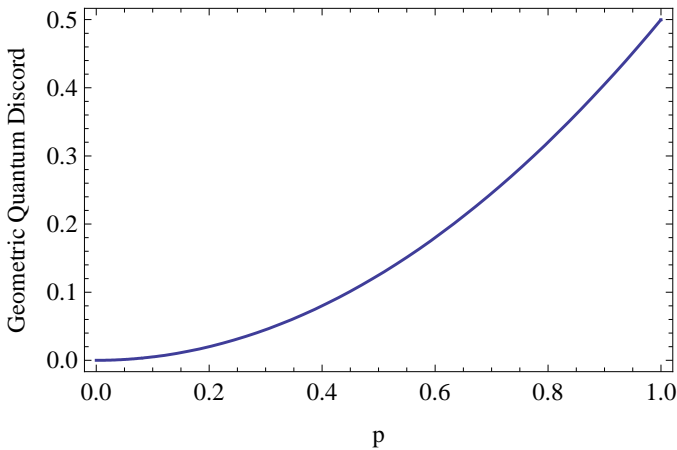}
\caption{Geometric Quantum Discord versus p. No dependence of
acceleration is observed and it reaches its maximum value when $ p=1
$.} \label{5}
\end{figure}

\indent In studying Quantum information of the particular state
introduced in this paper, we would like to investigate Geometric
Quantum Discord as an alternative definition of quantum
correlation. It is defined as the distance between the state being
studied and the one with zero QD. This is known as the
Hilbert-Schmidt distance \cite{DadicandLou}. In fact, Geometric QD
is the square norm of the Hilbert-Schmidt distance; hence, it is
called a 2-norm distance. Geometric QD could be quantitatively
expressed as
\begin{equation}\label{d27}
\mathcal{D}_G(\rho)=min||\rho-\rho'||.
\end{equation}
where, min is obtained over the measurements made. $
||\rho-\rho'||=Tr((\rho-\rho')^2) $ is the squared Hilbert-Schmidt
norm. $ \rho' $ is the state with zero QD. This is the state in
which the projective measurements $ \pi_i $ have been made on some
subsystems. This could be explicitly stated  in the bipartite
state as follows
\begin{equation}\label{d28}
\rho'=\sum_i (\pi_i \otimes 1)\rho (\pi_i \otimes 1),
\end{equation}
Finally, $ \mathcal{D}_G(\rho)=min_{\pi_i}||\rho-\rho'||=min_{\pi_i} Tr((\rho-\rho')^2)$.\\
\indent It has been shown that Geometric QD is degraded in
non-inertial frames  \cite{Brown}. Although the measure introduced
here and the one introduced in the previous subsection are
seemingly similar, they have certain basic differences. The latter
goes to zero in the infinite limit acceleration \cite{Brown} and
its maximum value tends to $ 0.5 $ rather than 1. We would like to
investigate this feature in the density matrix given in this
paper. Just similar to the case of QD, we would need to identify
the projective measures. The measurements which are to be made
effectively over a single qubit system are \cite{Brown}
\begin{equation}\label{d29}
\pi_\pm =\frac{1}{2}(I\pm\vec{x}.\vec{\sigma}).
\end{equation}
where, $I$ is the identity, $ \vec{x}=(x_1,x_2,x_3) $ is a unit
vector , $ x_1^2+x_2^2+x_3^2=1 $, and $
\vec{\sigma}=(\sigma_1,\sigma_2,\sigma_3)  $, and  $ \sigma $s are
pauli matrices. It has been clearly shown  that minimization
occurs when $ x_1=x_2=0, x_3=1 $  \cite{Datta}. Accordingly, $
\rho' $ for a bipartite system will take the following form:
\begin{eqnarray}\label{d30}
\rho' &=&\sum_{i=\pm} (\pi_i \otimes 1) \rho (\pi_i \otimes 1) \\
\nonumber &=&(1-e^{-2\pi\omega})^{2}
\sum_{n=0}^{\infty}e^{-2n\pi\omega}(n+1)\times\\ \nonumber
&(&\frac{1-p}{4}(|1\rangle_{+\uparrow A}|n+1\rangle_{-\uparrow
B}\langle1|_{+\uparrow A}\langle n+1|_{-\uparrow B} \\
\nonumber &+& |1\rangle_{+\downarrow A}|n+1\rangle_{-\downarrow
B}\langle 1|_{+\downarrow A}\langle n+1|_{-\downarrow B})\\
\nonumber &+&\frac{1+p}{4}(|1\rangle_{+\uparrow
A}|n+1\rangle_{-\downarrow
B}\langle 1|_{+\uparrow A}\langle n+1|_{-\downarrow B} \\
\nonumber &+& |1\rangle_{+\downarrow A}|n+1\rangle_{-\uparrow
B}\langle 1|_{+\downarrow A}\langle n+1|_{-\uparrow B})),
\end{eqnarray}
Using the basis $|1\rangle _{+\uparrow A}|n+1\rangle _{-\uparrow
B},\ \ |1\rangle _{+\uparrow A}|n+1\rangle _{-\downarrow B},\ \
|1\rangle _{+\downarrow A}|n+1\rangle _{-\uparrow B},\ \ |1\rangle
_{+\downarrow A}|n+1\rangle _{-\downarrow B}$, (\ref{d30}) could
be rewritten as follows
\begin{eqnarray}\label{d31}
\rho'&=&(1-e^{-2\pi\omega})^{2}\sum_{n=0}^{\infty}e^{-2n\pi\omega}\times \\
\nonumber&(n+1)& \left(
  \begin{array}{cccc}
    \frac{1-p}{4} & 0 & 0 & 0 \\
    0 & \frac{1+p}{4} & 0 & 0 \\
    0 & 0 & \frac{1+p}{4} & 0 \\
    0 & 0 & 0 & \frac{1-p}{4} \\
  \end{array}
\right),
\end{eqnarray}
which is the diagonal form of $ \rho $. Now, Geometric QD, which
is the trace norm of the difference between $\rho'$ in Eq.
(\ref{d31}) and $\rho$ from Eq. (\ref{d9}), is computed as in Eq.
(\ref{d32}) below
\begin{equation}\label{d32}
\mathcal{D}_G=\frac{p^2}{2}.
\end{equation}
Geometric QD is depicted versus p in Fig. \ref{5} . Again, no
effect of acceleration is seen and, as a consequence, it does not
make any difference in the infinite acceleration limit. Also, its
maximum value goes to $ 0.5 $ when $ p=1 $. This could also be
generalized to systems higher than the bipartite one. For a
general multipartite state, the global definition for Geometric QD
is \cite{Xu}
\begin{equation}\label{d33}
\mathcal{D}_G =min_{\lbrace \pi_{a} \rbrace } [Tr(\rho ^2)
-Tr(\rho'^{2})].
\end{equation}

As an extension, we could find $\rho'$ as a state with zero QD. It
is very interesting to see that  $\rho'$ in this case is the
diagonal form of $\rho$, too. Geometric QD would have the same
expression as Eq.(\ref{d32}). Again, acceleration does not have
any effect and the same properties hold as those for the bipartite
case
when $ p=1 $.\\
\indent In addition,  the Geometric QD as defined here has been
shown to run into some difficulties under local operations upon
the unmeasured subsystems \cite{piani}. A different definition is,
therefore,  introduced as the Geometric QD 1-norm which is the
only possible p-norm. This is known as Schatten 1-norm and is able
to consistently quantify non-classical correlations \cite{paula}.
It is defined as $ \mathcal{D}_G =minTr[\rho -\rho'] $, where $
\rho, \rho' $ are identified before. This is not in the interest
of this paper but it could be calculated and for the specific case
used here is equal to $p$.
\\

\section{Beyond single mode analysis}\label{e}
\indent Up until now, the results given are in the limit of the
single mode approximation. It is interesting to study whether or
not the entangled state introduced here preserves its specific
features beyond the single mode approximation. Generally, the
single mode approximation is not valid for all states. It  only
holds for some wave-packets related to the Minkowski ones if
fourier transforms are imposed. In this analysis, the entangled
state is quantified between a Minkowski and a Unruh mode instead
of the Rindler mode. This special mode is defined by its special
base and could be written as a linear combination of the Rindler
bases as follows\cite{Birrell}
\begin{eqnarray}\label{e1}
u_{\Omega,R}&=&\cosh(r_{\Omega})u_{\Omega,I}+\sinh(r_\Omega)u_{\Omega,II}^{\ast}, \nonumber \\
u_{\Omega,L}&=&\cosh(r_{\Omega})u_{\Omega,II}+\sinh(r_\Omega)u_{\Omega,I}^{\ast}.
\end{eqnarray}
In this equation, $u_{\Omega,R(L)}$'s are the Unruh right (left)
bases and $u_{\Omega,I(II)}$'s are the first (second) Rindler
bases, also $\tanh(r_{\Omega})=\exp(-\pi\Omega)$. Their
corresponding creation and annihilation operator could be
performed in a similar manner. Since the transformations between
Minkowski and Unruh modes do not mix, they could have the same
vacuum state \cite{Bruschi}
\begin{eqnarray}\label{e2}
|0_M\rangle&=&|0_{\Omega}\rangle = \prod_{\Omega}
|0_{\Omega}\rangle_{U},\nonumber \\
|0_{\Omega}\rangle_{U}&=&\sum_n
\frac{\tanh(r_{\Omega})^{n}}{\cosh(r_{\Omega})}
|n_{\Omega}\rangle_{I}|n_{\Omega}\rangle_{II}.
\end{eqnarray}
The other number states could be formed by implying the creation
Unruh operator on the vacuum state \cite{Bruschi}
\begin{eqnarray}\label{e3}
a^{\dagger}_{\Omega,U}|0_{\Omega}\rangle_U&=&\sum_{n=0}^{\infty}
\frac{\tanh(r_{\Omega})^{n}}{\cosh(r_{\Omega})}
\frac{\sqrt{n+1}}{\cosh(r_{\Omega})}|\phi^n_{\Omega}\rangle,
\\|\phi^n_{\Omega}\rangle&=&q_{L}|n_{\Omega}\rangle_{I}
|n+1_{\Omega}\rangle_{II}+q_{R}|n+1_{\Omega}\rangle_{I}|n_{\Omega}\rangle_{II}.\nonumber
\end{eqnarray}
where, $|q_R|^2 +|q_L|^2 =1$. The special case $q_R =1$, $q_L=0$
indicates the single mode approximation. Clearly, this special
choice breaks the symmetry between right and left wedges. For the
bipartite entangled state considered in this paper Eq.(\ref{d91}),
we need to define the state $|1_{\Omega}\rangle _{+\uparrow U}$
along the following lines:
\begin{eqnarray}\label{e4}
&|1_{\Omega}\rangle _{+\uparrow U}=(1-e^{-2\pi\Omega})
\sum_{n=0}^{\infty} e^{-2n\pi\Omega}\sqrt{(n+1)}&\\ \nonumber&(q_L
|n_{\Omega}\rangle^{I}_{+\uparrow}|(n+1)_{\Omega}\rangle^{II}_{-\uparrow}
+ q_R
|(n+1)_{\Omega}\rangle^{I}_{+\uparrow}|n_{\Omega}\rangle^{II}_{-\uparrow})&.
\end{eqnarray}
As  mentioned above, the density matrix should be computed  for
quantifying QE. By tracing over the second region, the result
would be obtained as the Alice-Bob density matrix; namely,
\begin{widetext}
\begin{eqnarray}\label{e5}
&\rho_{AB}&=\frac{1}{2}(1-e^{-2\pi\Omega})^2 \sum_{n=0}^{\infty}
e^{-2n\pi\Omega} (n+1)\times \\ \nonumber
&(&|q_L|^2(|1\rangle_{+\uparrow A}|n\rangle_ {-\downarrow
B}\langle 1|_{+\uparrow A}\langle n| _{-\downarrow B}+
|1\rangle_{+\downarrow A}|n\rangle_{-\downarrow B}\langle
1|_{+\uparrow A}\langle n|_{-\uparrow B}\\
\nonumber &+&|1\rangle_{+\uparrow A}|n\rangle_{-\uparrow
B}\langle1|_{+\downarrow A}\langle n|_{-\downarrow B} +
|1\rangle_{+\downarrow A}|n\rangle_{-\uparrow B}\langle
1|_{+\downarrow A}\langle n|_{-\uparrow B})\\ \nonumber &+&(
|q_R|^2(|1\rangle_{+\uparrow A}|n+1\rangle_{-\downarrow
B}\langle1|_{+\uparrow A}\langle n+1|_{-\downarrow B}+
|1\rangle_{+\downarrow A}|n+1\rangle_{-\downarrow B}\langle
1|_{+\uparrow A}\langle n+1|_{-\uparrow B}\\ \nonumber &+&
|1\rangle_{+\uparrow A}|n+1\rangle_{-\uparrow
B}\langle1|_{+\downarrow A}\langle n+1|_{-\downarrow B} +
|1\rangle_{+\downarrow A}|n+1\rangle_{-\uparrow B}\langle
1|_{+\downarrow A}\langle n+1|_{-\uparrow B})).
\end{eqnarray}
\end{widetext}
In the process of studying Negativity, the partial transpose of
the density matrix given in Eq. (\ref{e5}) should be taken. The
related eigenvalues are
\begin{eqnarray}\label{e6}
&\frac{1}{2}(1-e^{-2\pi\Omega})^2 \sum_{n=0}^{\infty}
e^{-2n\pi\Omega}\{|q_R|^2,\ |q_R|^2,\ |q_R|^2,&\ \nonumber \\
&-|q_R|^2, \ |q_L|^2,\ |q_L|^2,\ |q_L|^2,\ -|q_L|^2\}&.
\end{eqnarray}
Therefore, Negativity is  expressed by:
\begin{equation}\label{e7}
\log_{2} \frac{1}{2}(1-e^{-2\pi\Omega})^2 \sum_{n=0}^{\infty}
e^{-2n\pi\Omega}(4(|q_R|^2+|q_L|^2))=1.
\end{equation}
It is clearly seen that acceleration dependency disappears. Beyond
the single mode approximation,  we again find QE to be observer
independent. It is interesting that the dependence of $q_R$ and
$q_L$  also disappeared. By exchanging $q_R$ by $q_L$, the
Alice-AntiBob density matrix could be easily derived. The same
result is obtained for the Alice-AntiBob system. Going further, we
would like to investigate QD beyond the single mode approximation.
The approach adopted for calculating QD was introduced in Section
4. Defining the density matrix as in Eq.(26),  using Eq.(7) as
$|\psi\rangle$, and by tracing over the second (first) region, the
density matrix Eq.(\ref{d8}) for Alice-(Anti)Bob is derived. QD
for this special state beyond the single mode approximation is
obtained as follows
\begin{widetext}
\begin{eqnarray}\label{e8}
D(A;B)&=& \frac{1+3p}{4}|q_L|^2\log_2(\frac{1+3p}{4}|q_L|^2)
+\frac{1-p}{4}|q_L|^2\log_2(\frac{1-p}{4}|q_L|^2) \nonumber \\&+&
\frac{1+3p}{4}|q_R|^2\log_2(\frac{1+3p}{4}|q_R|^2)
+\frac{1-p}{4}|q_R|^2\log_2(\frac{1-p}{4}|q_R|^2) \nonumber \\&-&
\frac{1+3p}{4}|q_L|^2\log_2(\frac{1+p}{4}|q_L|^2) -
\frac{1-p}{4}|q_L|^2\log_2(\frac{1+p}{4}|q_L|^2) \nonumber \\&-&
\frac{1+3p}{4}|q_R|^2\log_2(\frac{1+p}{4}|q_R|^2) -
\frac{1-p}{4}|q_R|^2\log_2(\frac{1+p}{4}|q_R|^2).
\end{eqnarray}
\end{widetext}
Here, QD  only depends on $q_R$, $q_L$, and p. Using the
expression $|q_R|^2 +|q_L|^2 =1$, the above equation could be
further simplified to
\begin{eqnarray}
D(A;B)&=&\frac{(1-p)\log_2(1-p)-2(1+p)\log_2(1+p)}{\log_2 16} \nonumber\\
&+&\frac{(1+3p)\log_2(1+3p)}{\log_2 16} \nonumber
\\&=&\frac{1}{4}\log_2
\frac{(1+3p)^{1+3p}(1-p)^{1-p}}{(1+p)^{2(1+p)}}.
\end{eqnarray}
Therefore, it is independent of different values of $q_R$ and
$q_L$. Evidently, it is only p-dependent. If we factorize this
expression, we would exactly have  Eq.(\ref{d16}), which is
acceleration independent. It is observed that not only does
increment of acceleration have no impact on the Quantum features
for this specific helicity entangled state, but neither does
extending our approximation to  beyond the single mode  make any
difference to the value of Quantum Discord and Quantum
Entanglement. This is another interesting peculiarity of this
helicity entangled state which makes it a specific state in
Quantum Information Theory. Some other quantum features such as
Geometric Quantum Discord could be computed in this limit. The
tripartite helicity entangled state could also be a good candidate
for investigating QE and QD in the limit  beyond the single mode
approximation.

\section{Summary and conclusions}
\indent In summary, we investigated some quantum features of a
specific state which is entangled in the helicity part. It has
been shown that degradation of Quantum Entanglement for this
special case cancels when acceleration increases. Here, we
extended this feature to multipartite states. Particularly, we
showed that a tripartite entangled state yields the same result
and the degree of the entangled state remains unchanged. Moreover,
Quantum Discord as a measure of quantum correlation was studied
for the special bipartite and tripartite systems. A general
definition was given for multipartite systems. Using this global
definition, QD was computed for a specific tripartite system.
Acceleration was found to have no effect on this feature, either.
Another property  investigated in the present paper was the
special distance which tends toward zero in the infinite
acceleration limit. For the state considered here, however,
acceleration was found to disappear for the Geometric Quantum
Discord 2-norm. This last measure was seen to have potential
problems under local operations upon the unmeasured subsystems
\cite{piani}. The definition could have been possibly modified by
considering 1-norm of this distance, but it was not of interest in
this paper. These quantum features have been studied in the
analysis of beyond the single mode. A basic result which makes
this state an important state is that beyond the single mode
approximation, Negativity, as a measure of QE,  remains constant
in the value of a maximally entangled. Here, we have also computed
QD and noticed that this quantity which defines the degree of a
quantum correlation  also remains unchanged. By computing these
quantities, it is observed that quantum correlations for this
helicity entangled state is preserved in non-inertial frames in
both the single and beyond the single mode approximation.
Investigating QE, QD, and Geometric Quantum Discord for the
specific tripartite helicity entangled state which was introduced
here is suggested for further research in the limit of beyond the
single mode approximation.

{}
\end{document}